\documentclass[aip,graphicx]{revtex4-1}
\usepackage{graphicx}
\usepackage{amsmath}
\draft 

\begin{document}


\title{Amorphous Radial Frustration and Water-Like Anomalies in a Ramp-Shoulder Fluid} 



\author{Murilo Sodré Marques}
\email[]{murilo.sodre@ufob.edu.br}
\affiliation{Centro das Ciências Exatas e das Tecnologias, Universidade Federal do Oeste da Bahia, Rua Bertioga, 892, Morada Nobre, CEP 47810-059 Barreiras, BA, Brazil.}

\author{Lucas Axel Ramalho Santana}
\affiliation{Centro de Ciências Naturais e Humanas (CCNH), Universidade Federal do ABC, Avenida dos Estados, 5001, Bangu, CEP: 09210-580, Santo André-SP.}

\author{Gabriel San Rodrigues Ribeiro Câmara}
\affiliation{Instituto de Física, Universidade Federal de Uvberlândia, Av. João Naves de Ávila, 2121 – Bloco 1A, Bairro Santa Mônica, Uberlândia - MG, CEP 38408-100.}

\author{José Rafael Bordin}
\affiliation{Departamento de Física, Instituto de Física e Matemática,  Universidade Federal de Pelotas, Pelotas, Brazil.}
\affiliation{Fachbereich Physik, Universität Konstanz, Konstanz, Germany.}


\date{\today}

\begin{abstract}
\noindent\normalsize{We investigate the thermodynamic, structural, and dynamic behavior of a three-dimensional coarse-grained ramp-shoulder fluid derived from effective interactions between polymer-grafted nanoparticles. The interaction combines a softened repulsive ramp with a shallow attractive shoulder, stabilizing competing local organizations over a broad pressure interval. Molecular dynamics simulations reveal density, diffusion, and structural anomalies together with crystalline, amorphous, and fluid regions in the phase diagram. Unlike conventional isotropic core-softened fluids, the anomalous hierarchy becomes partially decoupled: the density anomaly extends beyond the structural anomaly, while the diffusion anomaly becomes closely connected to amorphization and shell migration processes. Analysis of radial distribution functions, excess entropy, translational and orientational order, and coordination-shell organization shows that the anomalies are not controlled solely by shell competition. Instead, they emerge from cooperative radial restructuring in a regime where radial correlations increase without the development of crystalline orientational order. The results indicate that the detailed shape of the softened interaction region strongly influences the structural pathways explored under compression, leading to a regime of amorphous radial frustration associated with anomalous diffusion and frustrated shell reorganization.}
\end{abstract}


\maketitle 

\section{Introduction}

The possibility of controlling collective behavior through effective interactions has become one of the central ideas in soft matter physics~\cite{Nagel2017,Manning2023,Barrat2023,Zhang2023}. In many colloidal and nanoparticle-based systems, the macroscopic response is not determined solely by chemical composition, but also by how particles organize at intermediate length scales~\cite{Likos2006,Jayaraman2013,Deblais2023}. Advances in coarse-grained modeling and computational methods have made it possible to investigate these effects using simplified interaction models that retain the essential physics while avoiding the complexity of fully atomistic descriptions~\cite{Likos2001,Huang2023,Noid2023,Maristany2025}.

Systems with softened or competing interaction ranges are especially interesting in this context because they can display unusual structural, thermodynamic, and dynamical properties. Water-like anomalies are among the best-known examples. In anomalous liquids, quantities such as density, diffusivity, or response functions vary nonmonotonically under compression or cooling, in contrast to the behavior expected for simple fluids \cite{gallo2016water,angell2000water,cheng2023origin}. Although water is the most familiar anomalous liquid, similar behavior has also been reported in colloids, liquid metals, silicon, phosphorus, and other complex fluids. This broader phenomenology suggests that anomalies are not necessarily tied to molecular specificity, but may instead arise from more general characteristics of the interaction potential.

Core-softened (CS) potentials became one of the main minimal models used to study these phenomena. In many CS systems, particles interact through two characteristic distances, and the competition between these local arrangements produces density, diffusion, and structural anomalies \cite{deOliveira2008,vilaseca2010softness,bordin2018distinct}. This interpretation established the idea that shell competition between two interaction scales constitutes the microscopic origin of anomalous behavior in isotropic systems.

At the same time, several studies indicate that this picture is incomplete. Comparisons between continuous and discontinuous shoulder potentials showed that the existence of two interaction scales alone is not sufficient to guarantee anomalies \cite{deOliveira2008b}. Likewise, the Gaussian-core fluid exhibits density, diffusion, and structural anomalies even though it does not possess a conventional two-scale interaction potential \cite{krekelberg2009}. In that system, anomalies emerge from cooperative structural reorganization involving correlations at different length scales rather than from direct competition between preferred interparticle distances. These results suggest that the microscopic origin of anomalies may depend not only on the presence of multiple scales, but also on how the system reorganizes structurally under compression.

This extensive perspective naturally connects anomalous fluids with problems involving frustration and amorphous organization. In water, for example, locally favored structures may develop an enhanced translational order while remaining frustrated against crystallization \cite{russo2014}. Related behavior has also been reported in square-shoulder and core-softened glass formers, where competing interaction ranges produce diffusion anomalies, slow dynamics, multiple glassy states, and complex relaxation behavior \cite{sperl2010,gnan2014}. These observations indicate that anomalies, amorphization, and frustrated ordering may be closely related manifestations of the same underlying structural competition.

An important ingredient in this problem is the detailed shape of the softened region of the interaction potential. Purely repulsive ramp-like interactions generally allow relatively smooth structural rearrangements during compression. However, softened shoulders containing shallow attractive regions can stabilize competing local organizations over broader thermodynamic intervals, favoring metastability and frustrated restructuring. Recent simulations of purely repulsive CS systems already suggested that water-like anomalies may emerge near solid--amorphous transformations even in the absence of a liquid--liquid phase transition \cite{bordin2023}. This raises the possibility that amorphization itself may play an active role in the anomalous behavior of soft-core fluids.

Polymer-grafted nanoparticles provide a natural realization of this scenario. The conformational degrees of freedom of the grafted chains allow the effective interaction landscape to be tuned through polymer architecture, flexibility, and grafting constraints. Depending on these conditions, the effective resulting interactions may show softened repulsive shoulders, shallow attractive regions, or competing interaction ranges capable of stabilizing different local organizations. Previous studies have shown that polymer-mediated interactions can generate ramp-shoulder potentials that display water-like anomalies and complex phase behavior in two dimensions \cite{marques2020a}. Whether the same mechanisms persist in three dimensions, where coordination and packing effects become substantially more complex, remains an open question.

In this work, we investigate the thermodynamic, structural, and dynamic behavior of a three-dimensional coarse-grained ramp-shoulder fluid derived from effective interactions between polymer-grafted nanoparticles. Our main goal is to understand how the shape of the softened interaction region influences the anomalous regime and the structural pathways explored under compression. We show that the anomalies are closely connected to cooperative radial restructuring processes that increase radial organization without stabilizing the crystalline orientational order. This mechanism generates a frustrated amorphous regime associated with shell migration, anomalous diffusion, and partial decoupling between thermodynamic and structural anomalies.

The remainder of this paper is organized as follows. In Section~\ref{modeling}, we introduce the models and simulation details. Section~\ref{res} presents and discusses the main results. Finally, we conclude with a summary of our findings and an outlook.

\section{Model and Simulation Details}
\label{modeling}

We investigate a coarse-grained model derived from effective interactions between polymer-grafted nanoparticles. Effective potentials were previously obtained from constrained polymer conformations using coarse-grained simulations combined with inversion approaches \cite{marques2020a,lafitte2014self}. Under constrained grafted-chain configurations, the resulting interaction displays a softened repulsive core and is hereafter denoted as \(U_{\mathrm{CS}}\).

\begin{figure}[h!]
    \centering
    \includegraphics[width=0.4\textwidth]{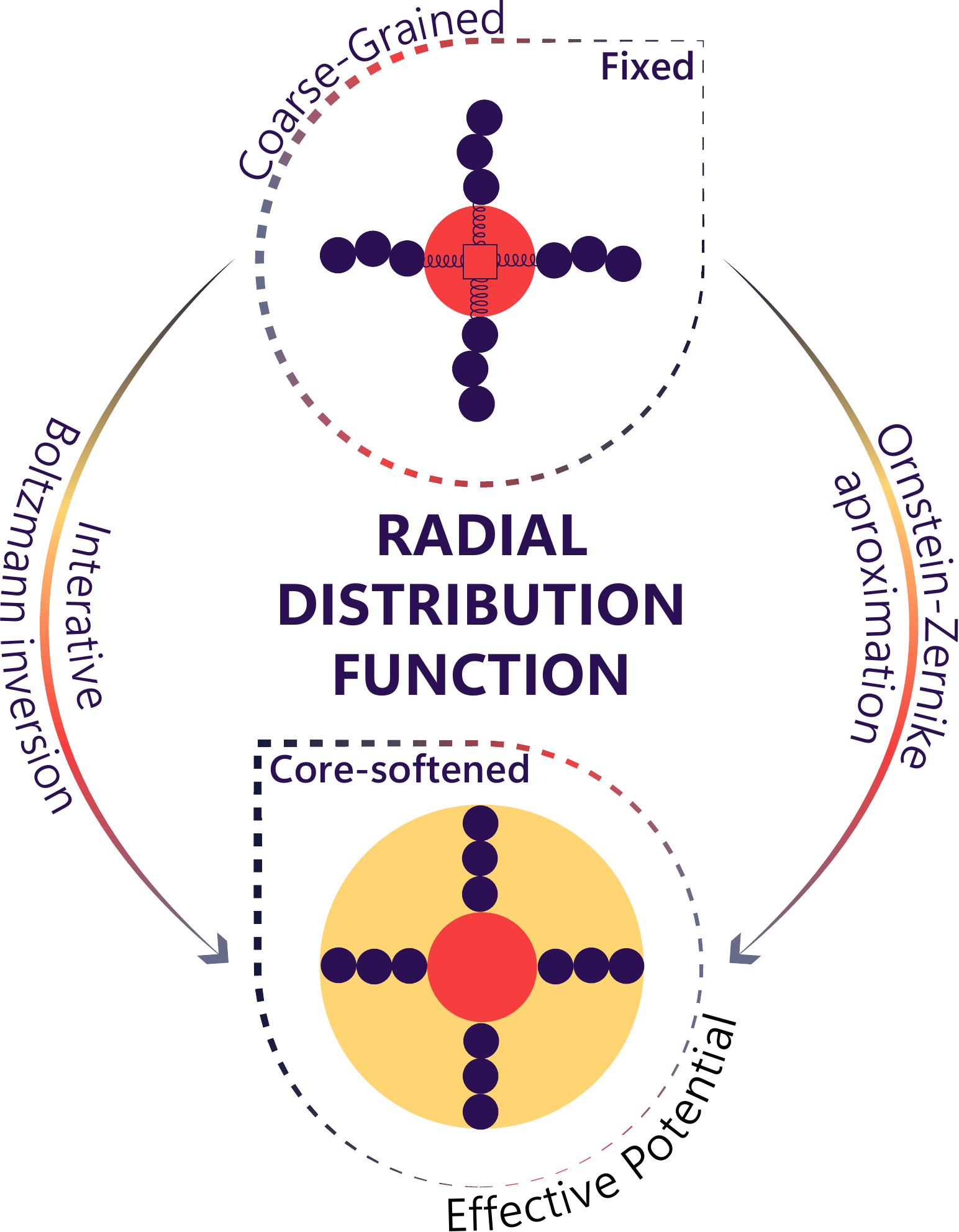}
   
\caption{Schematic representation of polymer-grafted nanoparticles and the corresponding effective interaction obtained from constrained polymer conformations, leading to a core-softened potential. The effective interaction is derived from the radial distribution function using inversion approaches such as iterative Boltzmann inversion and Ornstein--Zernike-based approximations.}
    \label{fig:diagrama}
\end{figure}

The effective interactions were extracted from radial distribution functions using iterative Boltzmann inversion and the Ornstein--Zernike equation with appropriate closures. Both approaches produced similar effective interactions. A schematic representation of the coarse-graining procedure is shown in Fig.~\ref{fig:diagrama}.

\begin{figure}[ht]
    \centering
    \includegraphics[width=0.65\textwidth]{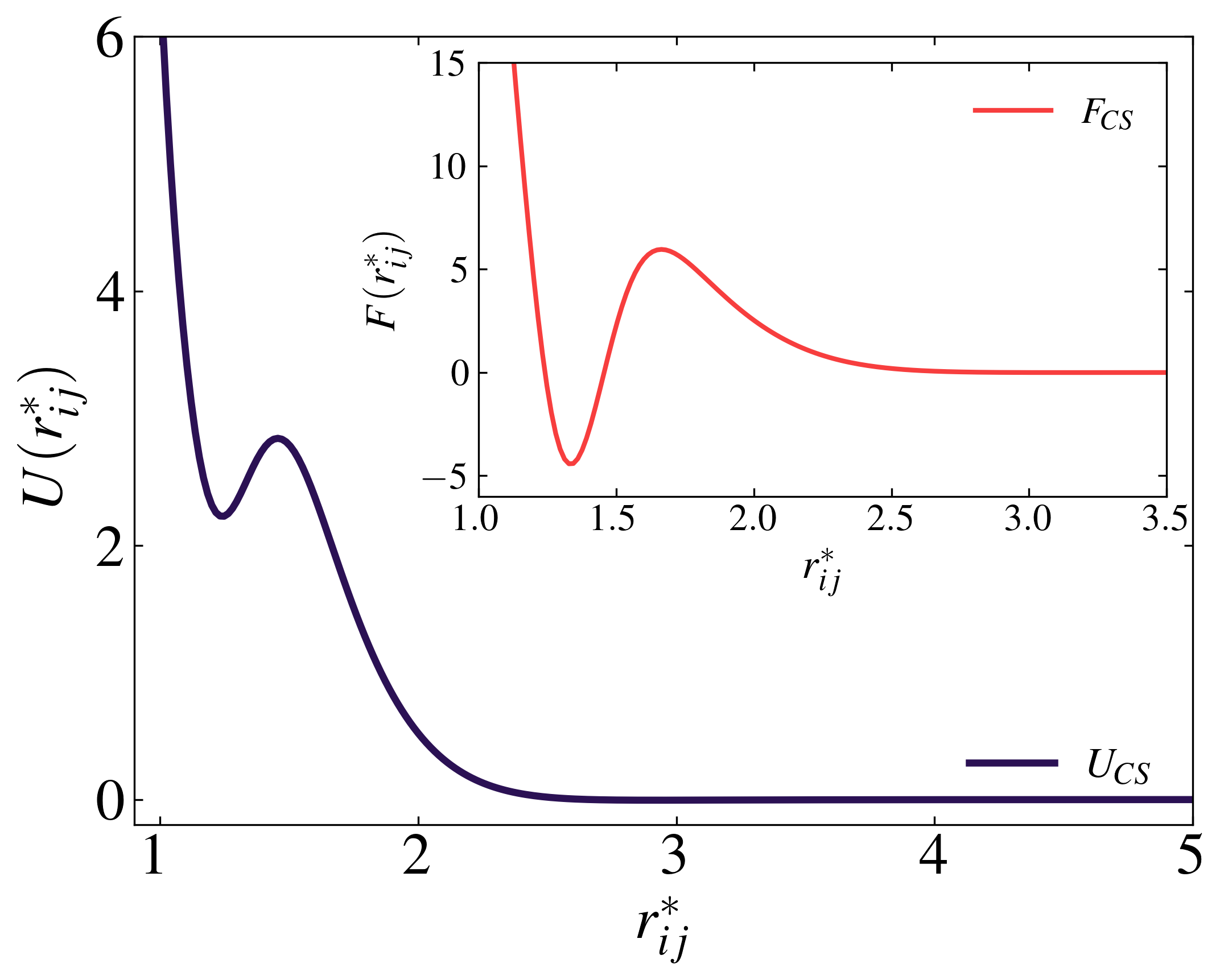}
     \caption{Effective interaction potential \(U_{\mathrm{CS}}\) and corresponding force (inset) as a function of the reduced interparticle distance \(r_{ij}^{\ast}\).}
    \label{fig:poten}
\end{figure}

The interaction potential is described by
\begin{equation}
U(r_{ij}) = 4\epsilon
\left[
\left(\frac{\sigma}{r_{ij}}\right)^{12}
-
\left(\frac{\sigma}{r_{ij}}\right)^{6}
\right]
+
\sum_{k=1}^{3}
h_k
\exp\left[
-
\left(
\frac{r_{ij}-c_k}{w_k}
\right)^2
\right],
\label{pot}
\end{equation}
where \(r_{ij}\) is the distance between the particles \(i\) and \(j\). The Lennard--Jones contribution describes the short-range interaction, while the Gaussian terms account for polymer-mediated effects. The parameters of the effective potential are listed in Table~\ref{tab1}. Figure~\ref{fig:poten} shows the interaction potential and the corresponding force.

\begin{table}[ht]
\centering
\caption{Parameters of the core-softened effective potential in reduced units.}
\resizebox{0.2\columnwidth}{!}{
\begin{tabular}{c c}
\hline
\multicolumn{2}{c}{\(U_{\mathrm{CS}}\)} \\
Parameter & Value \\
\hline
$h_1$ & $2.1895$  \\
$c_1$ & $0.8199$  \\
$w_1$ & $0.0420$  \\
$h_2$ & $9.6240$  \\
$c_2$ & $0.7947$  \\
$w_2$ & $0.7197$  \\
$h_3$ & $-3.8685$ \\
$c_3$ & $1.1684$  \\
$w_3$ & $0.2400$  \\
\hline
\end{tabular}
}
\label{tab1}
\end{table}

All simulations were performed in standard Lennard--Jones reduced units. The system consists of 1024 particles of diameter \(\sigma=1\) and mass \(m\) confined in a cubic simulation box with periodic boundary conditions. Molecular dynamics simulations were carried out using the LAMMPS package \cite{plimpton1995fast}. After equilibration in the NVE ensemble, production runs were performed in the NPT ensemble using a Nosé--Hoover thermostat and barostat. The integration time step was \(\delta t^{\ast}=0.001\).

Thermodynamic response functions were calculated from equilibrium fluctuations in the NPT ensemble. The specific heat at constant pressure is given by
\begin{equation}
c_P = \frac{\langle \delta H^2 \rangle}{k_B T^2},
\end{equation}
while the isothermal compressibility and thermal expansion coefficient are
\begin{equation}
\kappa_T = \frac{\langle \delta V^2 \rangle}{V k_B T},
\qquad
\alpha_P = \frac{\langle \delta V \delta H \rangle}{k_B T^2 V}.
\end{equation}

Structural properties were characterized through the radial distribution function \(g(r)\). The pair contribution to the excess entropy was computed as
\begin{equation}
S_2 = -2\pi\rho
\int_0^\infty
[g(r)\ln g(r) - g(r) + 1] r^2 dr,
\end{equation}
which provides the dominant contribution to the excess entropy in simple fluids \cite{sharma2006}.

To characterize long-range translational order, we also computed the cumulative two-body entropy,
\begin{equation}
C_{s2}(r) = - 2 \pi \rho \int_0^r
\left[
g(r') \ln g(r')
-g(r')
+1
\right]
r'^2 dr'.
\end{equation}

For fluids and amorphous phases, \(C_{s2}(r)\) converges at large distances, whereas crystalline phases display divergent behavior due to long-range translational order \cite{krekelberg2008,klumov2020}.

The translational order parameter was calculated as
\begin{equation}
\tau = \frac{1}{\xi_c}
\int_0^{\xi_c}
|g(\xi)-1| d\xi,
\end{equation}
while the coordination number was obtained from
\begin{equation}
N_{coord}(r) =
4 \pi \rho
\int_0^r
g(r') r'^2 dr',
\end{equation}
where the upper integration limit was chosen at the first or successive minima of \(g(r)\), depending on the number of coordination shells considered.

The susceptibility associated with translational order was computed as
\begin{equation}
\chi_{\tau} =
\left(
\dfrac{\partial \tau}{\partial P}
\right)_T,
\label{chitau}
\end{equation}
which measures the variation of translational order under compression.

The dynamic properties were analyzed using the diffusion coefficient \(D\), obtained from the Einstein relation :
\begin{equation}
D =
\dfrac{
\left< \Delta r^2(t) \right>
}
{2dt},
\end{equation}
where \(d\) is the dimensionality of the system.

Local orientational order was characterized using the Steinhardt bond-order parameter \(q_{\ell}\), defined as \cite{steinhardt1983,ten1996}
\begin{equation}
q_{\ell}(i) =
\dfrac{4 \pi}{2 \ell + 1}
\sum_{m= -\ell}^{+\ell}
\left |
q_{\ell m}(i)
\right |^2,
\end{equation}
where \(q_{\ell m}(i)\) is expressed in terms of spherical harmonics associated with the local bond orientations around particle \(i\).

\section{Results and Discussion}

\begin{figure}[h!]
    \centering
    \includegraphics[width=0.65\textwidth]{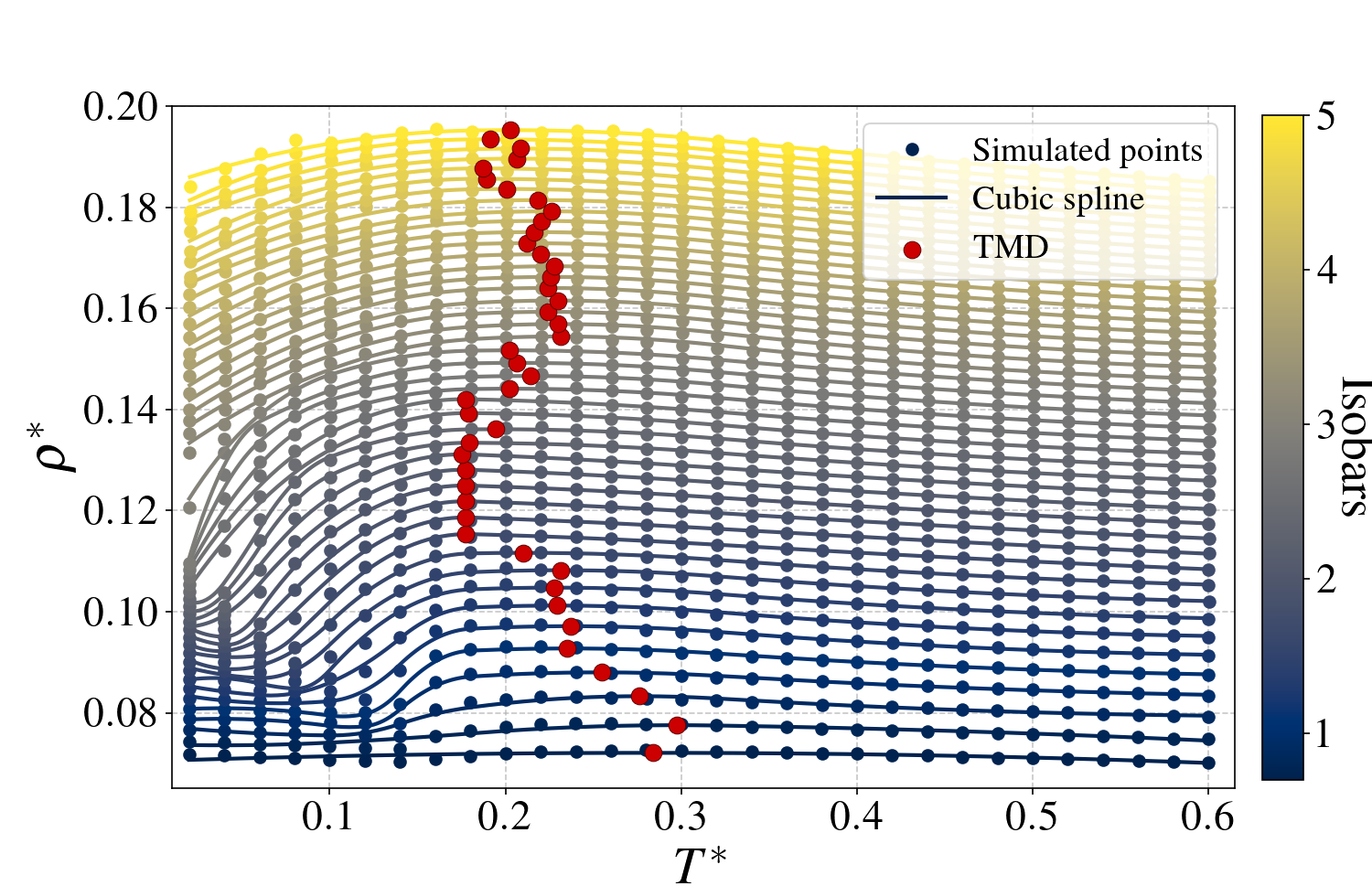}
    \caption{Temperature--density ($T^{\ast}$--$\rho^{\ast}$) diagram for the $U_{CS}$ potential. Points along the isobars indicate the temperatures of maximum density (TMD).}
    \label{rhoT_Uno}
\end{figure}

\begin{figure}[h!]
    \centering
    \includegraphics[width=0.65\textwidth]{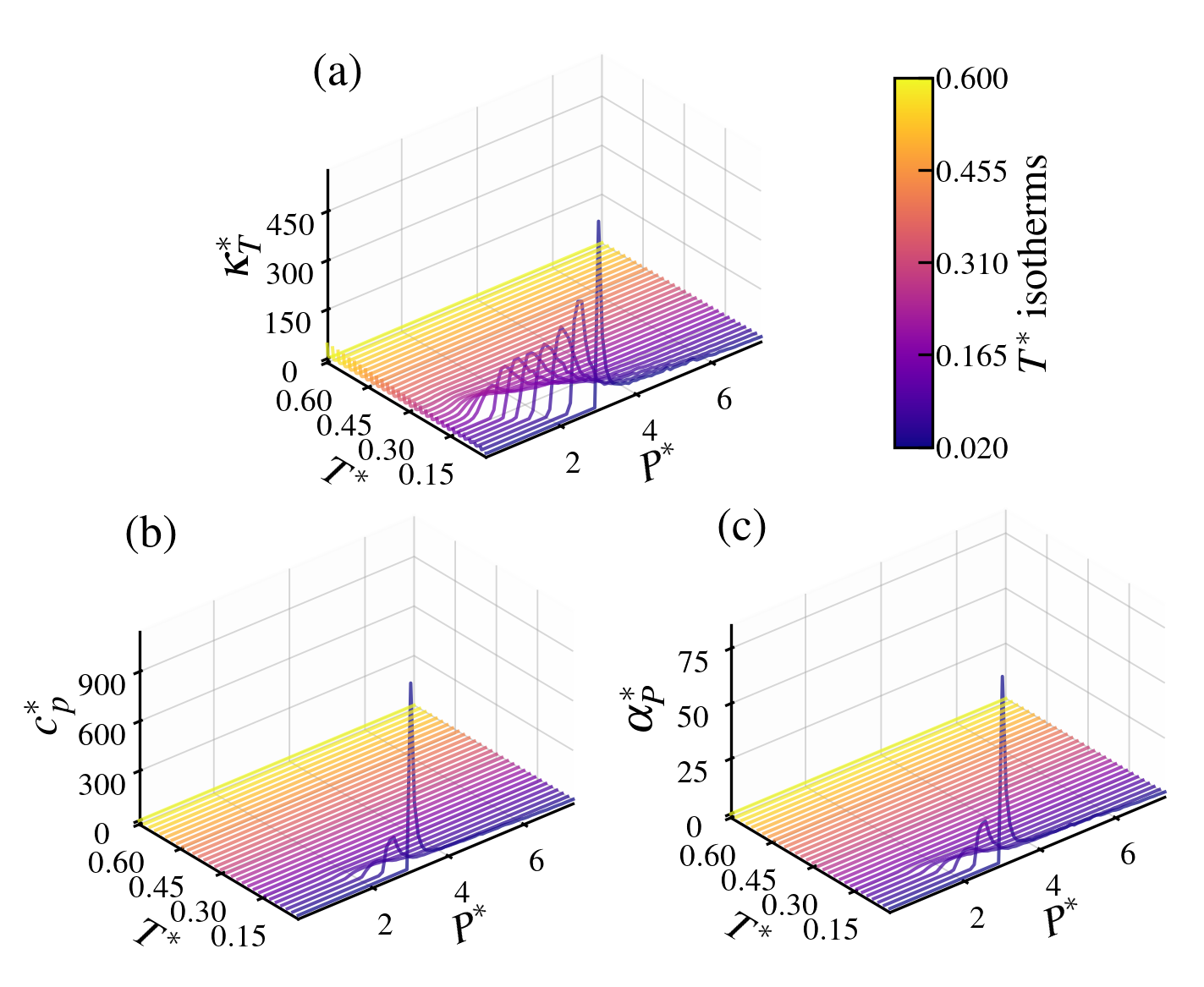}
    \caption{Thermodynamic response functions along isothermal paths for the $U_{CS}$ potential.}
    \label{thermoresponse}
\end{figure}

\begin{figure}[h!]
    \centering
    \includegraphics[scale=0.3]{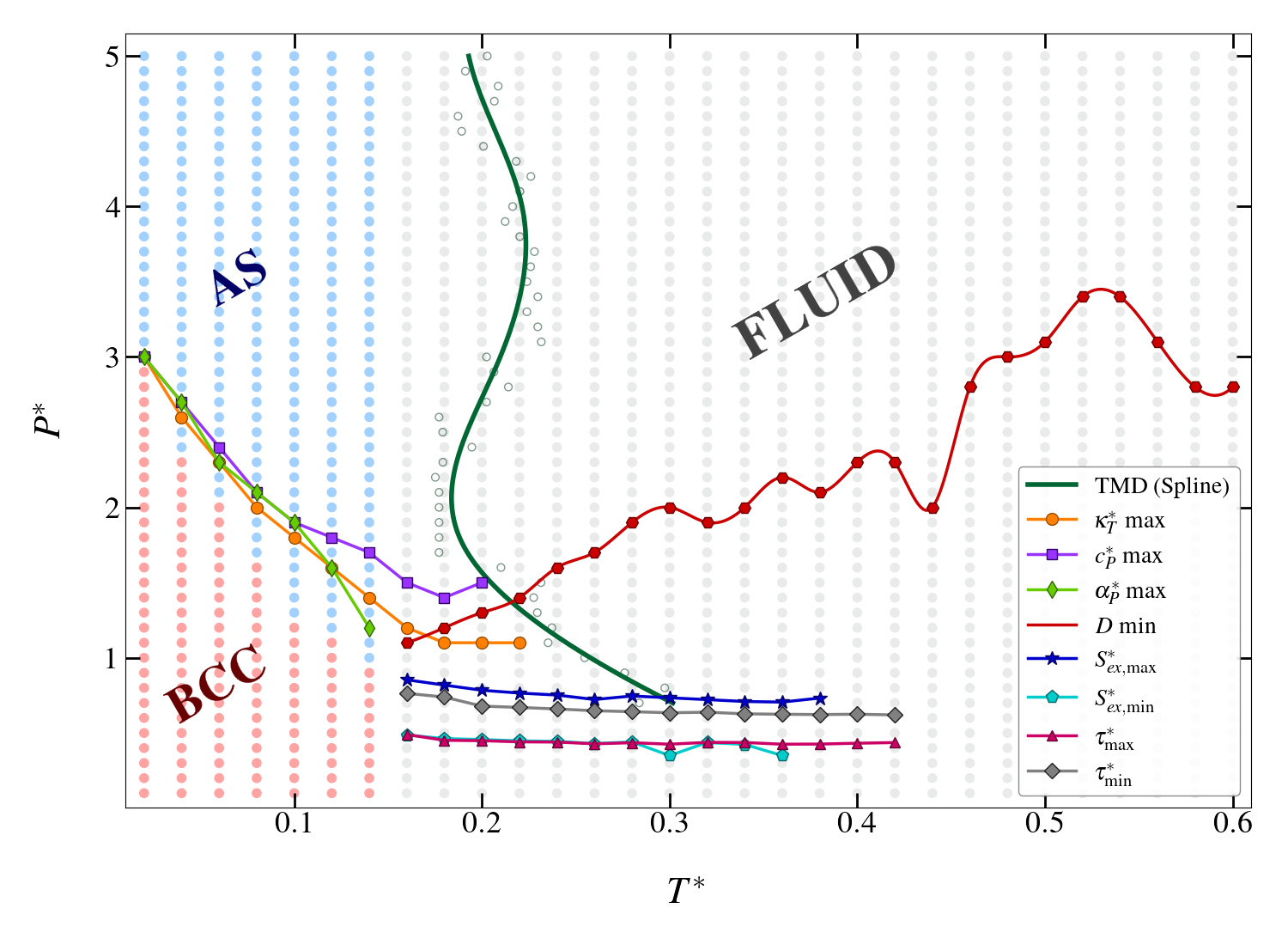}
    \caption{$P^{\ast}\times T^{\ast}$ phase diagram for the $U_{CS}$ potential. BCC denotes the body-centered cubic phase, AS the amorphous state, and Fluid the disordered fluid phase.}
    \label{PTUno}
\end{figure}

The $T^{\ast}\times\rho^{\ast}$ phase diagram for the $U_{CS}$ potential is shown in Fig.~\ref{rhoT_Uno}. The system exhibits a broad density-anomalous region, with the temperatures of maximum density (TMD) extending up to the highest investigated pressures. The TMD points were obtained along the isobars and combined with extrema of the thermodynamic response functions — namely the isothermal compressibility $\kappa_T$, thermal expansion coefficient $\alpha_P$, and specific heat $c_P$ (Fig.~\ref{thermoresponse}) — to construct the $P^{\ast}\times T^{\ast}$ phase diagram shown in Fig.~\ref{PTUno}. Structural characterization using PTM analysis in OVITO further allowed the identification of the BCC crystalline phase, the amorphous state, and the fluid region.

At first glance, the phase behavior resembles that commonly observed in isotropic core-softened systems, where the competition between interaction distances produces thermodynamic and dynamic anomalies. However, structural and dynamical properties reveal important differences. In conventional water-like CS fluids, the anomalous hierarchy usually follows the sequence structural anomaly $\rightarrow$ diffusion anomaly $\rightarrow$ TMD \cite{deOliveira2008,franzese2010,fomin2013}. Here, the anomalies become partially decoupled. The TMD line extends toward considerably higher pressures than the structural anomaly, while the dynamic anomaly appears to be closely connected to amorphization and frustrated restructuring.

\begin{figure}[h!]
    \centering
    \includegraphics[width=0.8\textwidth]{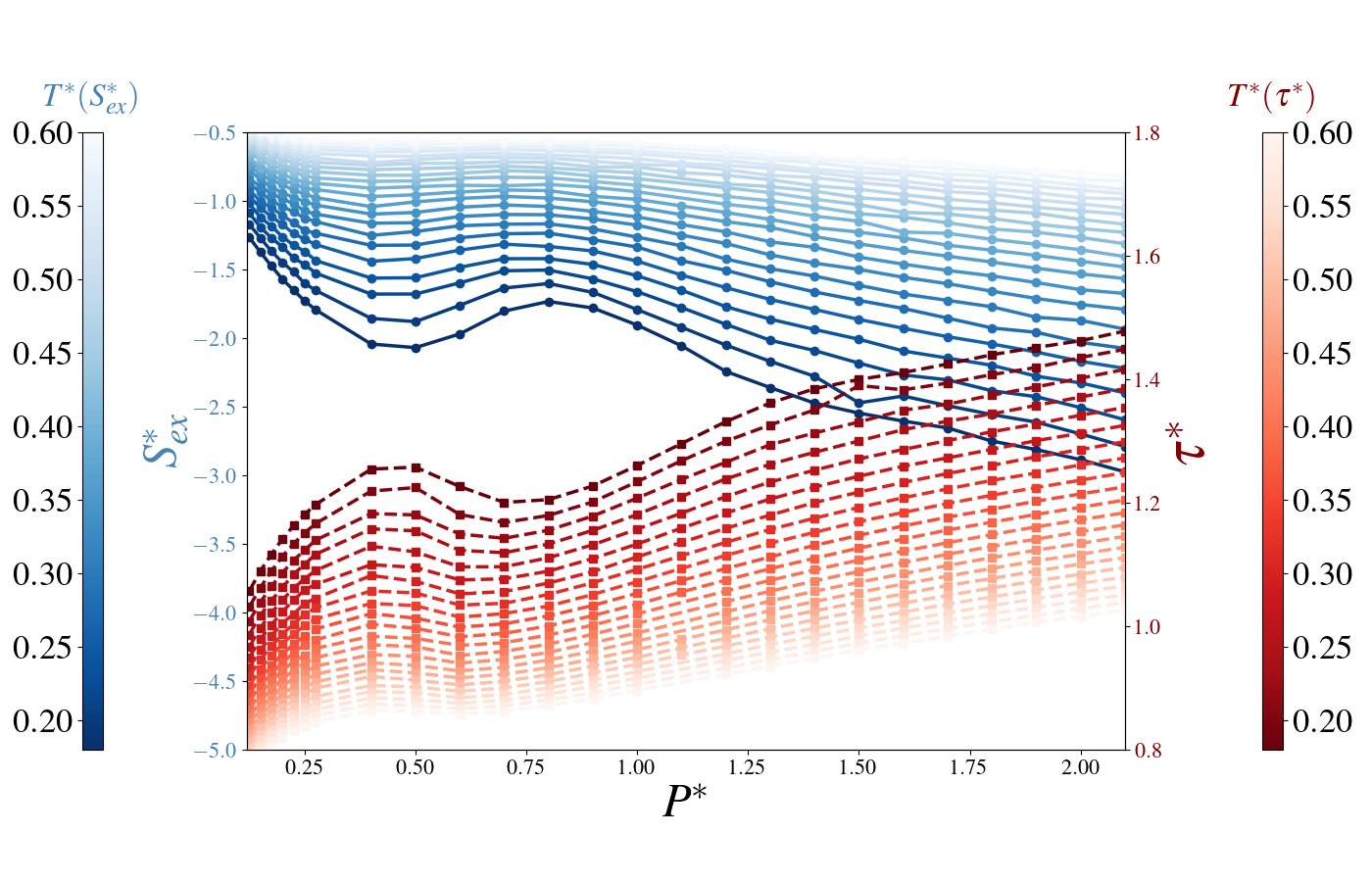}
    \caption{Pair excess entropy $s_2$ and translational order parameter $\tau$ for the $U_{CS}$ potential.}
    \label{sextau_Uno}
\end{figure}

The translational order parameter $\tau$ and the pair excess entropy $s_2$, shown in Fig.~\ref{sextau_Uno}, help clarify this behavior. For temperatures above approximately $T^{\ast}\approx0.18$, both quantities vary smoothly with pressure, consistent with a fluid phase. Within the interval $0.50\lesssim P^{\ast}\lesssim0.80$, however, the system displays anomalous structural behavior: $\tau$ increases under compression while $s_2$ becomes less negative. Outside this region, the fluid recovers the expected monotonic behavior of simple liquids, namely increasing translational order accompanied by decreasing excess entropy.

\begin{figure}[h!]
    \centering
    \includegraphics[width=0.65\textwidth]{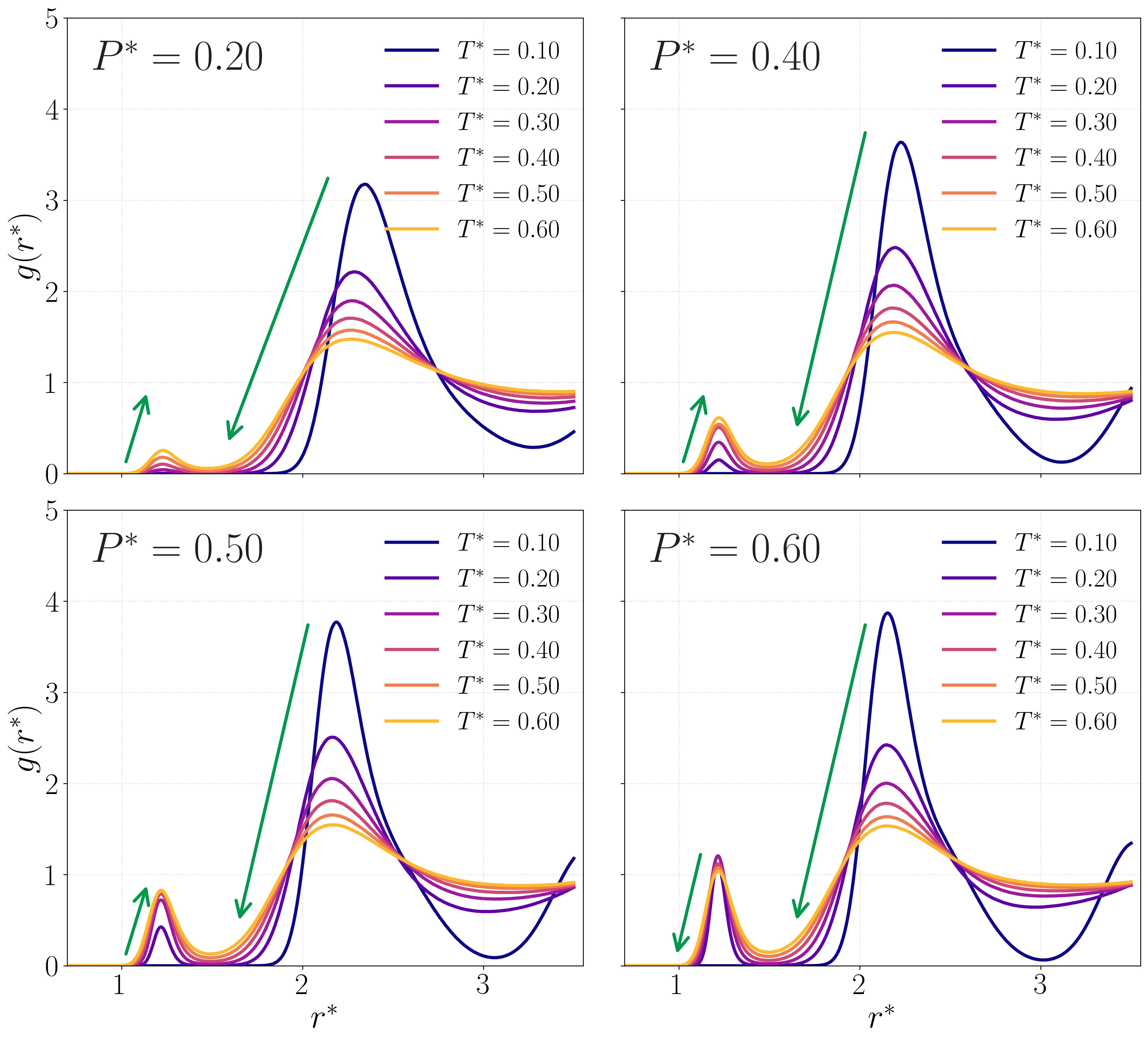}
    \caption{Radial distribution functions illustrating thermally induced competition between the characteristic length scales of the $U_{CS}$ potential.}
    \label{tempeffects}
\end{figure}

To examine the microscopic origin of this restructuring, the radial distribution functions were analyzed along different thermodynamic paths. At low pressures, the competition between scales is primarily thermally driven. In the NPT ensemble, this condition can be expressed as \cite{deOliveira2008,vilaseca2010softness}

\begin{equation}
\Pi_{12}=
\left(\frac{\partial g(r)}{\partial T}\right)_{P,r_1}
\left(\frac{\partial g(r)}{\partial T}\right)_{P,r_2}
<0,
\label{termaleq}
\end{equation}

where $r_1$ and $r_2$ correspond to the shorter and longer characteristic scales of the interaction potential. Figure~\ref{tempeffects} shows that increasing temperature transfers the population toward the shorter scale, opposite to the behavior expected for simple liquids. Thermal fluctuations destabilize the open local arrangement associated with $r_2$ and favor denser local configurations associated with $r_1$.

At higher pressures, shell competition becomes predominantly compression driven. In this case,

\begin{equation}
\Pi_{12}=
\left(\frac{\partial g(r)}{\partial P}\right)_{T,r_1}
\left(\frac{\partial g(r)}{\partial P}\right)_{T,r_2}
<0.
\label{presscomp}
\end{equation}

\begin{figure}[h!]
    \centering
    \includegraphics[width=0.65\textwidth]{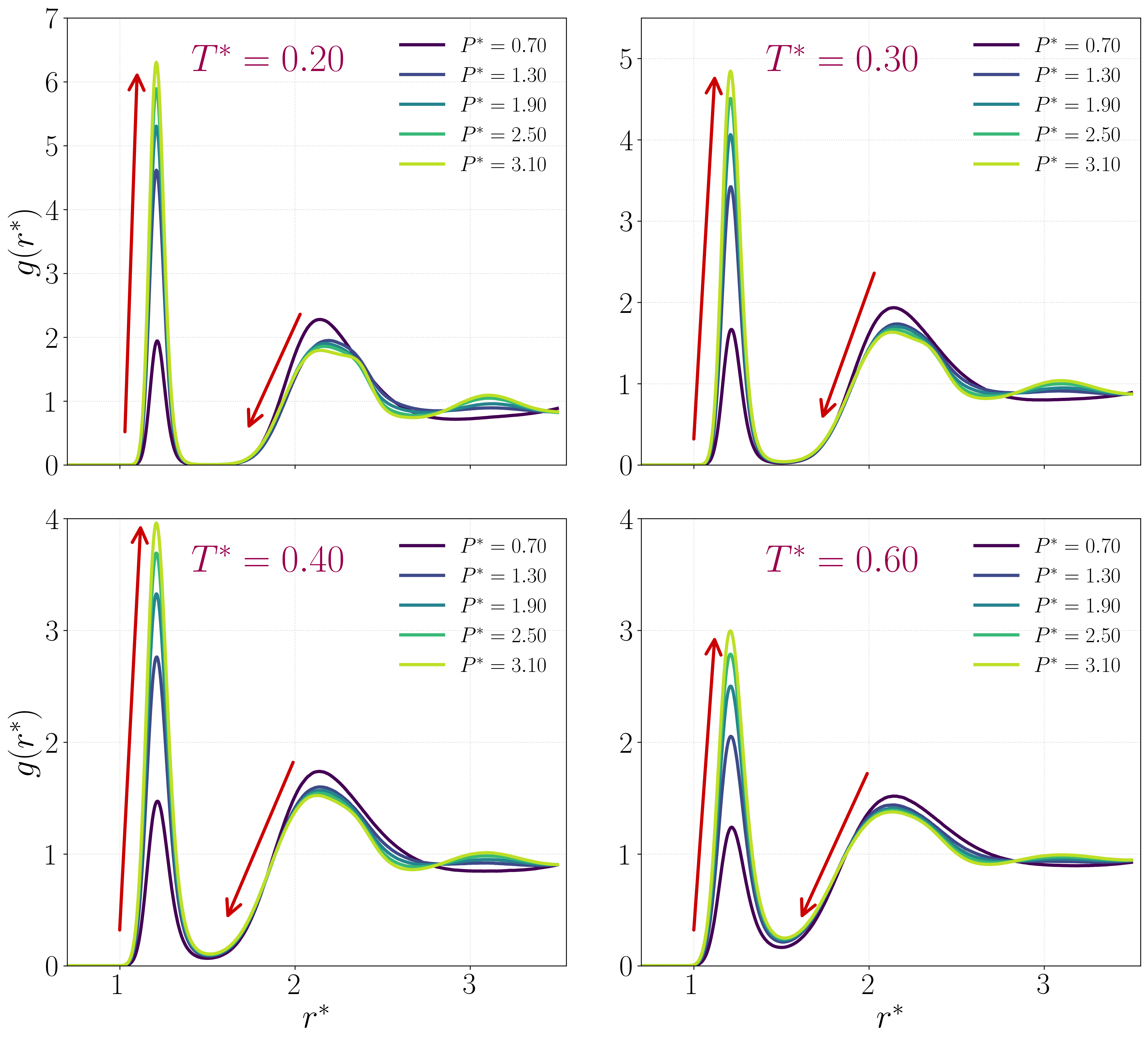}
    \caption{Radial distribution functions showing pressure-induced competition between the characteristic length scales of the $U_{CS}$ potential.}
    \label{presseffects_Uno}
\end{figure}

Figure~\ref{presseffects_Uno} shows that, for approximately $0.50<P^{\ast}<3.60$, compression transfers the population from the outer shell to the inner shell. Similar shell migration processes were previously observed in other core-softened systems \cite{marques2020a}. Here, however, the restructuring extends over a broader pressure interval and becomes strongly connected to amorphous organization.

An important property of the present model is the shape of the softened region of the interaction potential. Unlike purely repulsive ramp-like models, this $U_{CS}$ interaction contains a shallow attractive shoulder. As a consequence, particles occupying the outer coordination shell are not only geometrically stabilized, but also energetically favored over a finite pressure interval. Compression, therefore, does not simply eliminate the open-shell arrangement. Instead, particles continuously redistribute between competing local organizations, promoting frustrated restructuring and amorphous configurations. This behavior is consistent with previous comparisons between continuous and discontinuous shoulder potentials, where anomalies were shown to depend strongly on the detailed form of the softened region and on the resulting excess-entropy behavior \cite{deOliveira2008b, evy2013}.

\begin{figure}[h!]
    \centering
    \includegraphics[width=0.65\textwidth]{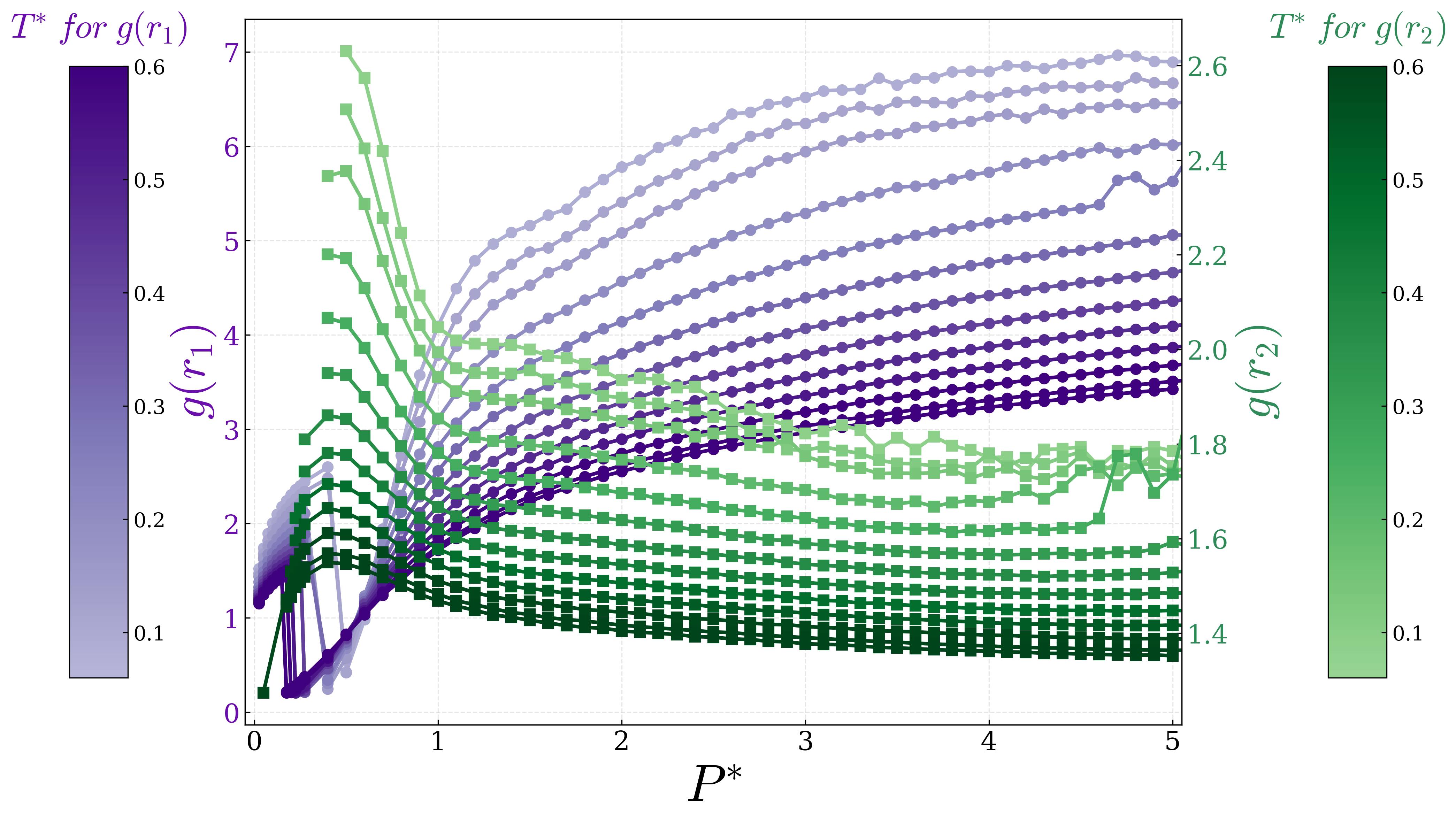}
    \caption{Peak values of $g(r_1)$ and $g(r_2)$ as a function of pressure, illustrating the crossover between thermally and pressure-driven shell competition in the $U_{CS}$ potential.}
    \label{tempffects_Uno}
\end{figure}

The evolution of the RDF peak heights further illustrates this process. As shown in Fig.~\ref{tempffects_Uno}, both $g(r_1)$ and $g(r_2)$ increase monotonically at low pressures, consistent with thermally dominated shell occupation. Around $P^{\ast}\approx0.50$, the outer-shell population begins to decrease while the inner-shell population increases. This crossover coincides with the onset of the anomalous structural and dynamic behavior.

Interestingly, condition $\Pi_{12}<0$ remains satisfied even outside the anomalous region. The existence of competing shell populations alone is, therefore, insufficient to generate anomalies in this system. Similar behavior was reported for the Gaussian-core fluid, where anomalies also arise from cooperative structural reorganization rather than from explicit competition between two preferred interaction distances \cite{krekelberg2009}. In the present model, the anomalies appear only when shell migration becomes sufficiently cooperative to produce large-scale radial restructuring without stabilizing the orientational order of the crystalline. The attractive and shallow shoulder plays an important role in this mechanism. Because the outer-shell arrangement remains energetically favorable over a broad pressure interval, the system cannot continuously reorganize into a compact crystalline structure during compression. Instead, competing local organizations persist over an extended thermodynamic region, increasing structural frustration and favoring amorphous restructuring.

\begin{figure}[h!]
    \centering
    \includegraphics[width=0.65\textwidth]{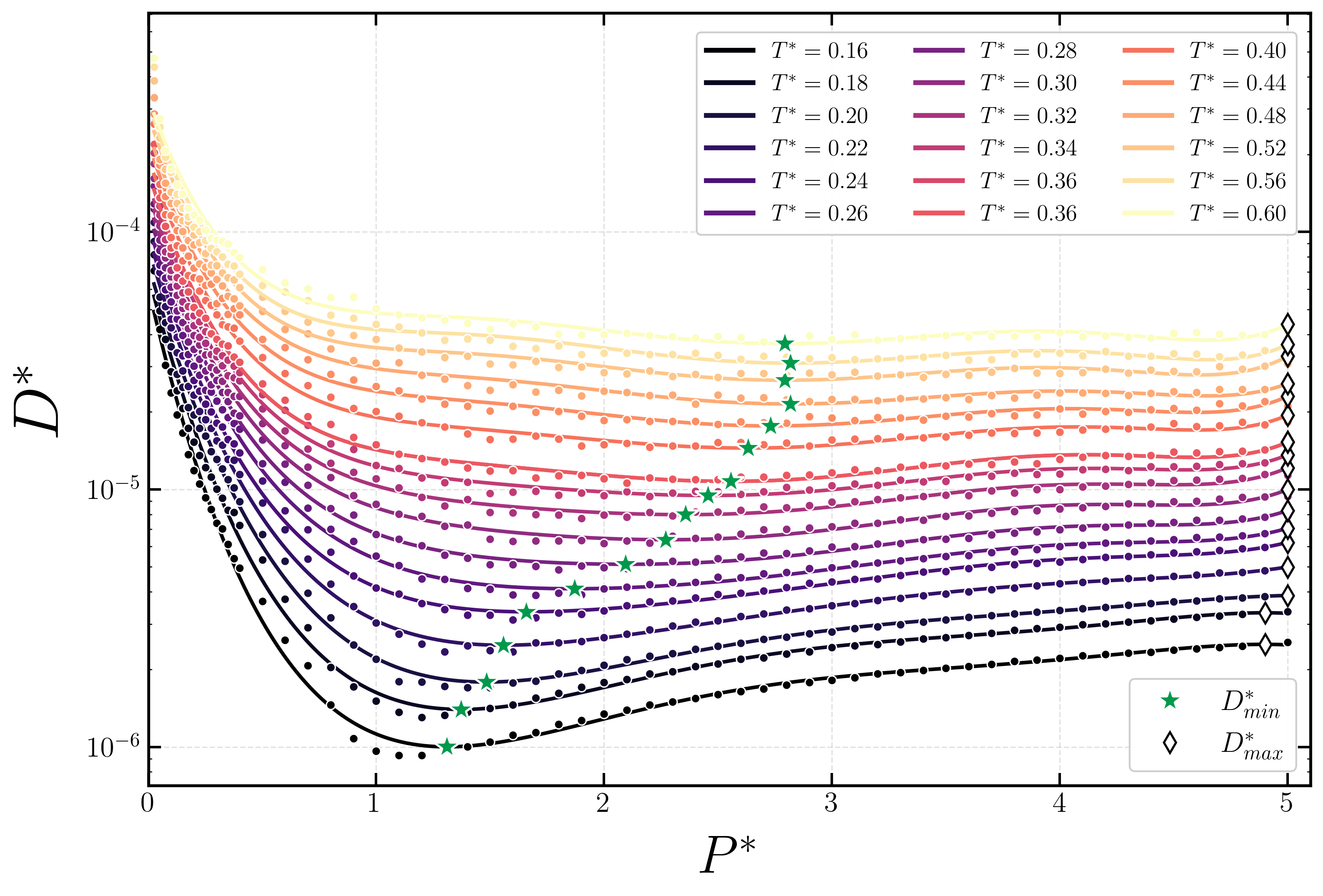}
    \caption{Diffusion coefficient along selected isotherms for the $U_{CS}$ potential.}
    \label{diff_Uno}
\end{figure}

The dynamic behavior follows the same trend. Figure~\ref{diff_Uno} shows that the diffusion coefficient develops a minimum along the isotherms for $T^{\ast}\gtrsim0.16$. Initially, compression suppresses mobility because the particles become increasingly trapped in transient local cages. Further compression destabilizes the open-shell arrangement and promotes migration toward the shorter scale, partially restoring mobility and generating the anomalous regime, where

    \begin{equation}
        \left(\frac{\partial D}{\partial P}\right)_T>0.
    \end{equation}

The diffusion minimum appears close to the pressure interval where the low-temperature BCC phase loses stability, and amorphous configurations become dominant. This connection between anomalous dynamics and amorphization differs from the conventional picture of isotropic CS systems, where anomalous diffusion is usually interpreted mainly in terms of local shell competition. A similar relation between amorphization and water-like anomalies was previously observed in a purely repulsive ramp-like CS fluid \cite{bordin2023}.

\begin{figure}[h!]
    \centering
    \includegraphics[width=0.65\textwidth]{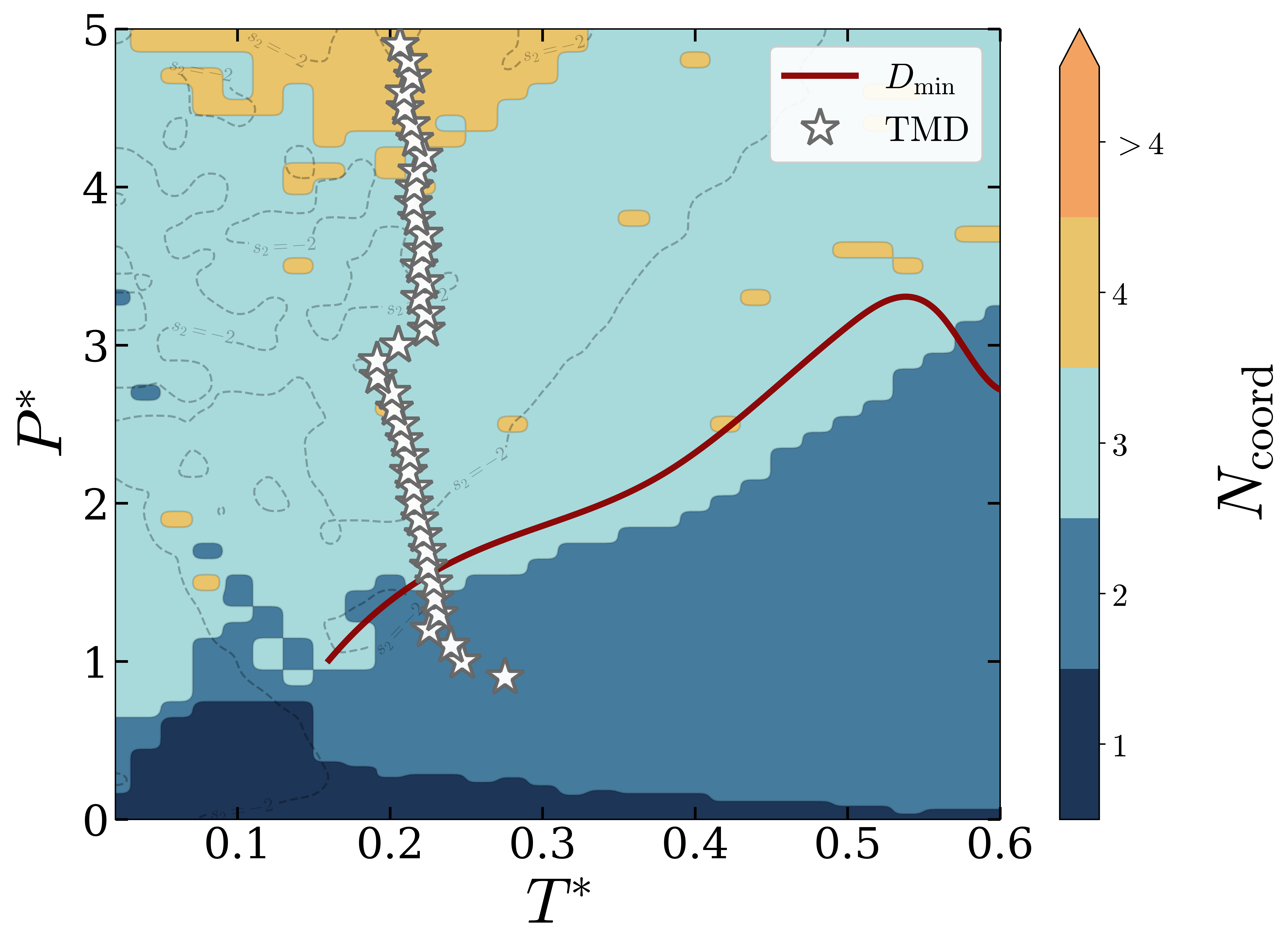}
    \caption{$P^{\ast}\times T^{\ast}$ structural map showing the evolution of radial correlations and anomalous regions. Dashed lines indicate the locus where $s_2=-2$.}
    \label{PTmap}
\end{figure}

To characterize the structural changes associated with this process, we constructed the map shown in Fig.~\ref{PTmap} based on the effective coordination-shell number $N_{coord}$. Three distinct regimes emerge. At low pressures, where $N_{coord}\approx1$--$2$, the system behaves as a weakly correlated fluid with short-range organization. At intermediate pressures, where $N_{coord}\approx2$--$3$, structural correlations extend over larger distances, and shell competition becomes stronger. This region coincides with the diffusion anomaly and with the extrema of $\tau$ and $s_2$. At higher pressures, where $N_{coord}\approx4$, the fluid develops stronger translational organization and more persistent radial correlations.

The TMD line does not coincide with the structural boundaries defined by $N_{coord}$. This partial decoupling between thermodynamic and structural anomalies differs from the hierarchy commonly reported for water-like CS models and indicates that the density anomaly is not controlled solely by the monotonic growth of translational order.

\begin{figure}[h!]
    \centering
    \includegraphics[width=0.65\textwidth]{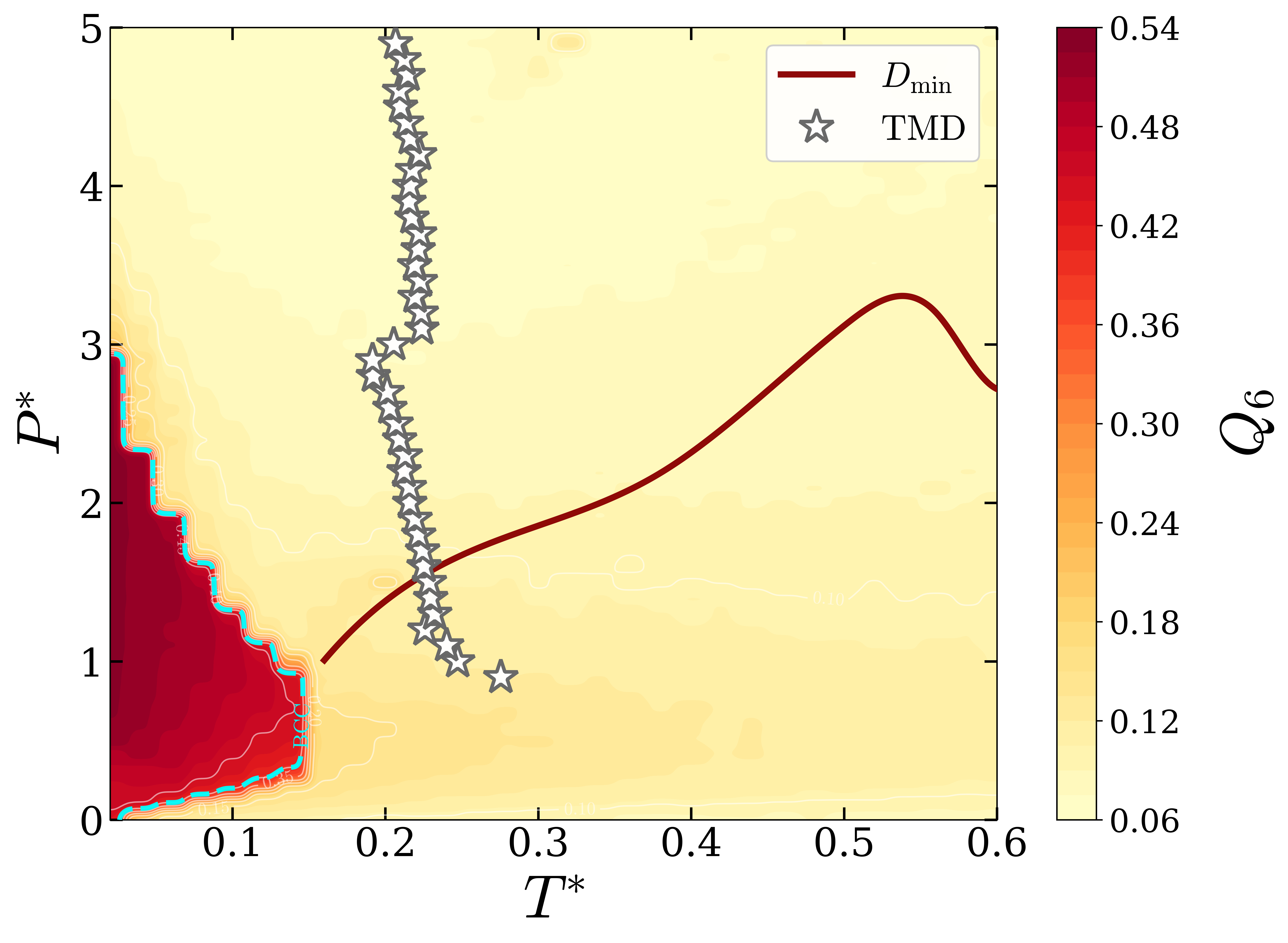}
    \caption{$P^{\ast}\times T^{\ast}$ map of the orientational order parameter $Q_6$ for the $U_{CS}$ potential.}
    \label{Q6-PTUcs}
\end{figure}

Since both $N_{coord}$ and $s_2$ are cumulative quantities derived from $g(r)$, their correlation indicates that the anomalies are fundamentally associated with radial restructuring processes. The orientational order parameter $Q_6$, shown in Fig.~\ref{Q6-PTUcs}, demonstrates that this restructuring is not accompanied by significant orientational crystallization.

At low temperatures, compression initially drives the system toward local BCC order. Throughout the anomalous region itself, however, $Q_6$ remains relatively low even when $\tau$ and $s_2$ exhibit pronounced anomalies. Radial organization therefore increases while orientational order remains suppressed. This behavior differs from several previously studied core-softened systems, where anomalous regions are frequently associated with precursor crystalline order or tetrahedral local organization. It also resembles the locally favoured structure scenario proposed for water, in which enhanced translational ordering coexists with frustration against crystallization \cite{russo2014}.

The anomalous region therefore corresponds to a regime of amorphous radial frustration. Compression enhances radial organization by continuously redistributing particles between the two characteristic scales of the interaction potential. This behavior is reflected in the evolution of $N_{coord}$, the extrema of $\tau$ and $s_2$, the minima in the diffusion coefficient, and the pressure dependence of the RDF peak populations. At the same time, crystalline orientational order remains suppressed, as indicated by the low values of $Q_6$ throughout the anomalous region.

Rather than resolving the competition between scales through crystallization, the system remains trapped in a frustrated amorphous state where radial order grows without long-range orientational organization. The anomalous regime may therefore be interpreted as a broad avoided structural transition between competing local fluid arrangements associated with the two characteristic interaction distances.

\begin{figure}[h!]
    \centering
    \includegraphics[width=0.65\textwidth]{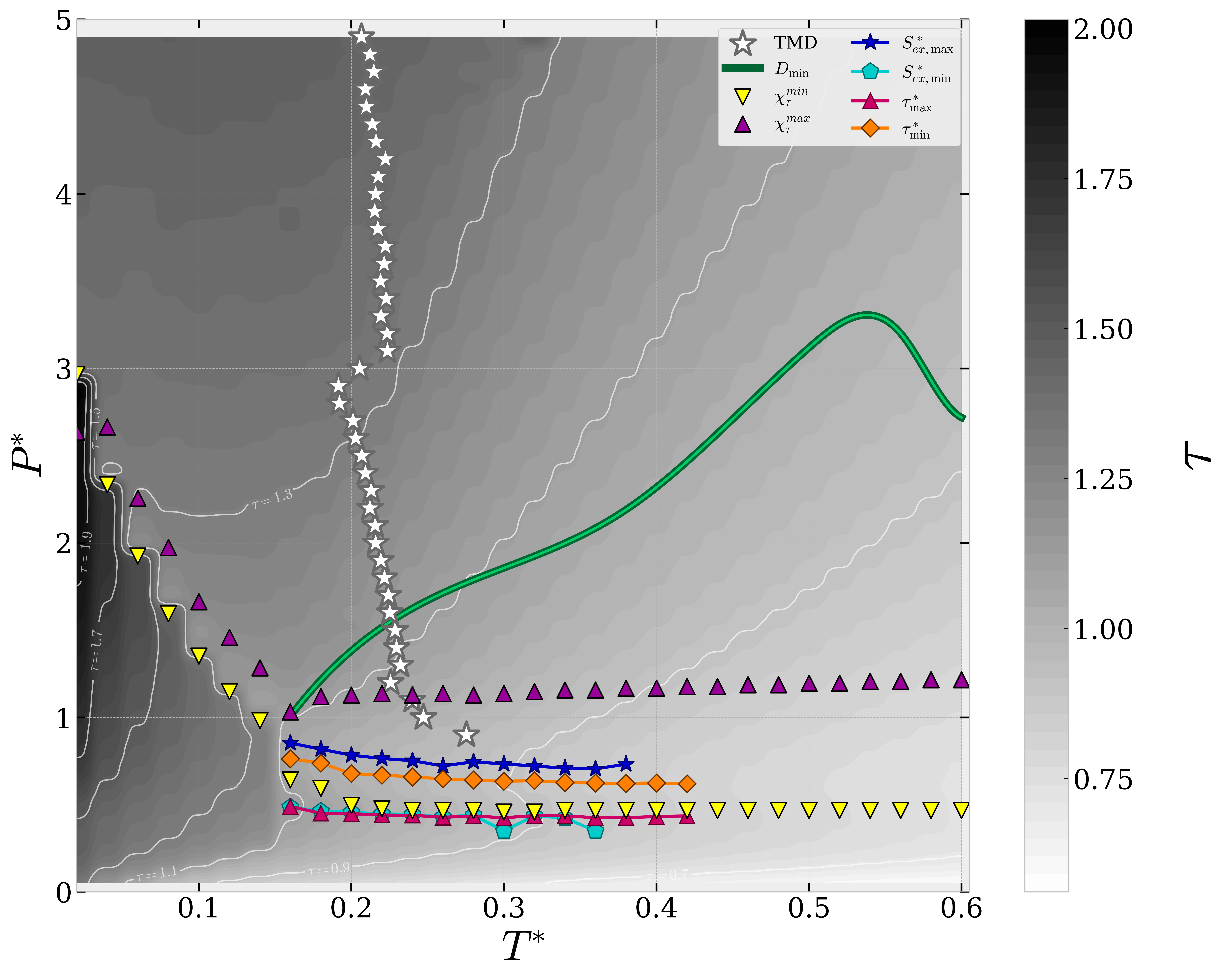}
    \caption{$P^{\ast}\times T^{\ast}$ map showing the extrema of the translational-order susceptibility $\chi_{\tau}$.}
    \label{tauPTmap}
\end{figure}

The susceptibility of the translational order parameter, $\chi_{\tau}$, reinforces this picture. Figure~\ref{tauPTmap} shows that pronounced minima in $\chi_{\tau}$ appear near the pressure interval where the BCC phase disappears, and the amorphous state becomes dominant. At higher temperatures, the extrema of $\chi_{\tau}$ persist even in the fully fluid regime and progressively shift toward lower pressures, remaining approximately within

\[
0.40\lesssim P^{\ast}\lesssim0.85,
\]

which coincides with the anomalous regions of $\tau$, $s_2$, and $N_{coord}\sim2$. Similar connections between shell competition, amorphization, and anomalous dynamics were previously reported for square-shoulder systems displaying diffusion anomalies and complex glassy dynamics \cite{sperl2010,gnan2014}.

In general, the present results show that the anomalies of the fluid $U_{CS}$ are closely related to the cooperative radial restructuring in a regime in which the radial correlations increase without the development of the orientational order of the crystal. The softened attractive shoulder stabilizes competing local organizations over a broad pressure interval, leading to shell migration, frustrated amorphous restructuring, and partial decoupling between thermodynamic and structural anomalies. In this sense, the anomalous behavior of the system cannot be understood solely in terms of the existence of two interaction scales, but rather through the structural pathways explored during compression.

\section{Conclusions}

In this work, we investigated the thermodynamic, structural, and dynamic behavior of a three-dimensional coarse-grained ramp-shoulder fluid derived from effective interactions between polymer-grafted nanoparticles. The system displays density, diffusion, and structural anomalies together with crystalline, amorphous, and fluid regions in the phase diagram. Although these anomalies resemble those commonly observed in isotropic core-softened fluids, the underlying structural mechanism differs from the conventional picture based solely on shell competition between two interaction distances.

The anomalous regime originates from a cooperative radial restructuring process associated with the shape of the ramp-shoulder interaction. The shallow attractive shoulder stabilizes open-shell local arrangements over a broad pressure interval, preventing the fluid from reorganizing smoothly into a compact crystalline structure during compression. Instead, particles continuously redistribute between competing coordination shells, leading to frustrated amorphous restructuring.

The structural analysis based on radial distribution functions, excess entropy, translational order, coordination-shell organization, and orientational order shows that the anomalous regime corresponds to a state of amorphous radial frustration. In this region, radial correlations increase considerably while orientational crystalline order remains weak. The anomalies therefore emerge not simply from the existence of two interaction scales, but from the structural pathways explored during compression.

Another important result is the partial decoupling of the conventional anomalous hierarchy. The density anomaly extends beyond the structural anomaly, whereas the diffusion anomaly becomes closely connected to amorphization and shell migration processes. Moreover, the condition associated with shell competition remains satisfied even outside the anomalous region, indicating that competing shell populations alone are insufficient to generate anomalies.

Finally, our results show that the detailed shape of softened interaction regions can strongly influence how anomalous fluids reorganize under compression. More generally, they suggest that amorphous radial frustration may provide a useful framework for understanding anomalous behavior in coarse-grained soft-core systems and related softened-interaction fluids.


%
%

%

\begin{acknowledgments}
We thank the financial support from the National Council for Scientific and Technological Development CNPq (441728/2023-5). J.R.B. thank CNPq (304958/2022-0), Research Support Foundation of the State of Rio Grande do Sul FAPERGS grant 25/2551-0002918-0, and the Coordination for the Improvement of Higher Education Personnel - Brazil (CAPES)/Alexander von Humboldt Foundation for financial support through a research fellowship.
\end{acknowledgments}

\bibliography{aipsamp.bib}

\end{document}